\documentstyle[aps,preprint,tighten,floats,psfig]{revtex}
\begin{document}
\draft
\preprint{\vbox{\it Submitted to Phys. Rev. D \hfill\rm CU-NPL-1154\\
                \null \hfill ADP-97-44/T272
               } }

\title{\bf Light Hadron Spectroscopy on Coarse Lattices \\
with \pmb{$O(a^2)$} Mean-Field Improved Actions}

\author{Frank X. Lee\footnote{
E-mail:~fxlee@sammy.colorado.edu ~$\bullet$~ Tel:
303~492--3659 ~$\bullet$~ Fax: 303~492--7486 \hfill\break
\null\quad\quad WWW:~http://stripe.colorado.edu/$^\sim$flee/}}
\address{Nuclear Physics Laboratory, Department of Physics,
University of Colorado, \\Boulder, CO 80309-0446}
\address{TRIUMF, 4004 Wesbrook Mall Vancouver, BC, Canada V6T 2A3}
\author{Derek B. Leinweber\footnote{
E-mail:~dleinweb@physics.adelaide.edu.au ~$\bullet$~ Tel:
+61~8~8303--3548 ~$\bullet$~ Fax: +61~8~8303--3551 \hfill\break
\null\quad\quad WWW:~http://www.physics.adelaide.edu.au/$^\sim$dleinweb/}}
\address{Special Research Center for the Subatomic Structure of Matter,
University of Adelaide 5005}
\address{Department of Physics and Mathematical Physics,
University of Adelaide 5005, Australia}
\address{Department of Physics, University of Washington, Seattle, WA
98195, USA}
\maketitle

\begin{abstract}
The masses and dispersions of light hadrons are calculated in lattice
QCD using an $O(a^2)$ tadpole-improved gluon action and an $O(a^2)$
tadpole-improved next-nearest-neighbor fermion action originally
proposed by Hamber and Wu.  Two lattices of constant volume with
lattice spacings of approximately 0.40 fm and 0.24 fm are considered.
The results reveal some scaling violations at the coarser lattice
spacing on the order of 5\%.  At the finer lattice spacing, the
calculated mass ratios reproduce state-of-the-art results using
unimproved actions.  Good dispersion and rotational invariance up to
momenta of $p\,a\simeq 1$ are also found.  The relative merit of
alternative choices for improvement operators is assessed through
close comparisons with other plaquette-based tadpole-improved actions.
\end{abstract}

\vspace{1cm}
\parskip=2mm

\section{Introduction}

Lattice discretization of the continuum QCD action introduces errors
at finite lattice spacing $a$.  The standard Wilson gauge action has
$O(a^2)$ errors and the standard Wilson fermion action has $O(a)$
errors.  Simulations using these actions have shown that lattice
spacings of 0.1 fm or less and lattice volumes of $24^4$ or larger are
needed in order to hold systematic errors to the 10\% level.  Such
simulations are major undertakings and require enormous computing
power to extract even the most basic of hadronic observables, the
hadron masses (see~\cite{Butler94} for an example and~\cite{Yos97} for
a recent review).

During the past few years, considerable efforts have been devoted to
improving lattice actions.  The idea is to reduce or remove the
discretization errors from the actions so that they have better
continuum-like behavior.  At the same time, errors due to the lattice
regularization are accounted for through the renormalization of the
coefficients multiplying the improvement operators.  The hope is the
use of improved actions will allow one to simulate efficiently and
accurately on coarse lattices, such that computer resources may be
redirected to the simulation of QCD rather than quenched QCD.
Moreover, one may turn the focus of investigation towards quantities
of experimental interest.

The way to develop improved discretizations of continuum actions is
not unique.  There are numerous programs in this endeavor having their
foundation in one of two formulations.  One is based on lattice
perturbation theory which is used to derive the renormalized
coefficients through a mean-field reordering of the perturbative
expansion, as recently reviewed by Lepage~\cite{Lep97}.  The other is
based on renormalization group theory to construct fixed-point (or
perfect) actions, as recently reviewed by Hasenfratz~\cite{Has97}.  

In the pure gauge sector, the $O(a^2)$ tadpole-improved action
of~\cite{Alf95} leads to dramatic improvement in the static potential
and glueball masses~\cite{Mor97} up to lattice spacings of 0.4 fm.
Excellent scaling and topological properties have been demonstrated in
fixed-point actions~\cite{DeG96}.  In the light quark sector, hadron
spectroscopy has been investigated with a variety of improved actions
including the $O(a)$-improved SW action~\cite{Col97}, the
$O(a^2)$-improved D234 action~\cite{Alf95a} and its
variants~\cite{Lep97,Alf96}, and the D$\chi34$ action of Hamber and Wu
\cite{Ham83,Egu84,Fie96,Lee97} considered here.  Fixed-point fermion
actions have been studied in free field theory~\cite{Bie97,DeG97a} and
in simulations~\cite{DeG97}.

In the present work, we report further calculations of the light
hadron spectrum using the $O(a^2)$ tadpole-improved gluon action
of~\cite{Alf95} and the D$\chi34$ action proposed by Hamber and
Wu~\cite{Ham83} many years ago.  This action is an $O(a^2)$
next-nearest-neighbor fermion action with tadpole-improved estimates
of the coupling renormalizations.  This action is selected primarily
due to its simplicity.  The cost of simulating it is about a factor of
two as compared to standard Wilson fermions.  Our goal is to study its
feasibility as an alternative action to SW which has the clover term,
or to D234 which has both next-nearest-neighbor couplings and the
clover term.

This work extends previous studies of the D$\chi34$ action
\cite{Fie96,Lee97} through finer lattice spacings, improved
statistics, and the simulation of additional observables.  In
particular, we examine dispersion relations and test the rotational
symmetry of both the gauge and the fermion actions.  Hadron mass
ratios are calculated for a wide variety of hadrons including
hyperons.  To explore scaling violations, we consider two coarse
lattices of approximately fixed physical volume: $6^3\times 12$ at a
lattice spacing of 0.40 fm and $10^3\times 16$ at 0.24 fm.

\section{Improved Lattice Actions}

\subsection{The Gauge Action}

The improved gauge action employed in this investigation is given
by~\cite{Alf95}:
\begin{equation}
S_G=\beta\sum_{\mbox{pl}}{1\over 3} \mbox{Re Tr}(1-U_{\mbox{pl}})
-{\beta\over 20 u^2_0}\sum_{\mbox{rt}}{1\over 3} 
\mbox{Re Tr}(1-U_{\mbox{rt}}) \, .
\end{equation}
The second term removes the $O(a^2)$ errors at tree level.
Perturbative corrections are estimated to be of the order of
2-3\%~\cite{Alf95}.  Here, $U_{\mbox{rt}}$ denotes the rectangular 1x2
plaquettes.  $u_0$ is the tadpole factor that largely corrects for the
large quantum renormalization of the links $U_\mu(x)=\exp(i \, g\,
\int_x^{x+a \widehat\mu} A(y) \cdot dy )$.  In this calculation we
use the mean plaquette to estimate $u_0$,
\begin{equation}
u_0\equiv \left({1\over 3} \mbox{Re Tr}
\langle U_{\mbox{pl}}\rangle\right)^{1/4} \, ,
\label{u0}
\end{equation}
and will focus our evaluation of lattice action improvement on other
plaquette-based improved actions.  $u_0$ is determined
self-consistently in the simulation.  An alternative choice is to use
the mean link in Landau gauge $u_0\equiv \langle{1\over 3}
\mbox{Tr}U_\mu\rangle_{LG}$ ~\cite{Alf96}.

\subsection{The D$\chi34$ Fermion Action}

The improvement program of Sheikoleslami and Wohlert (SW)~\cite{She85}
provides a systematic approach to the improvement of lattice fermion
actions.  However, the on-shell improvement program leaves some
freedom in the relative values of the coefficients of the improvement
operators.  In this investigation, we consider a specific case of the
general class of D234 actions \cite{Alf96} in which the improvement
parameters are tuned to remove the second-order
chiral-symmetry-breaking Wilson term.  This fermion action may be
written
\begin{equation}
M_{\rm{D}\chi34} = m_q + \gamma \cdot \nabla + {1 \over 6} \sum_\mu
\left ( -a^2 \nabla_\mu \Delta_\mu + b \, a^3 \Delta_\mu^2 \right ) \,
\label{Dchi34}
\end{equation}
where
\begin{equation}
\nabla_\mu \, \psi(x) = {1 \over 2a} \left [
U_\mu(x) \, \psi(x+\mu) - U_\mu^\dagger(x-\mu) \, \psi(x-\mu) \right ]
\, ,
\end{equation}
and
\begin{equation}
\Delta_\mu \, \psi(x) = {1 \over a^2} \left [
U_\mu(x) \, \psi(x+\mu) + U_\mu^\dagger(x-\mu) \, \psi(x-\mu) - 2 \,
\psi(x) \right ] \, .
\end{equation}
The second-order term of the D234 action
\begin{equation}
\sum_\mu \Delta_\mu + { 1 \over 2} \, \sigma \cdot F
\end{equation}
breaks chiral symmetry and does not appear in the D$\chi34$ action.
However, the fourth-order term of (\ref{Dchi34}) also breaks chiral
symmetry and provides for the removal of the fermion doublers.

Explicit evaluation of (\ref{Dchi34}) combined with a
Wilson-fermion-style field renormalization factor of $(3 \, \kappa /
2)$ discloses the following simple fermion action\cite{Ham83,Egu84}:
\begin{eqnarray}
S_F &=& \sum_{x}\bar{\psi}(x)\psi(x) \nonumber \\
& & - \kappa \sum_{x,\mu} \left[
\bar{\psi}(x) \, (b-\gamma_\mu) \, {U_\mu(x)\over u_0} \, \psi(x+\mu)
+\bar{\psi}(x+\mu) \, (b+\gamma_\mu) \, {U_\mu^\dagger(x)\over u_0} \,
\psi(x) \right.
\nonumber \\ & &\qquad
+\bar{\psi}(x) \, (-{1 \over 4}\, b + {1 \over 8} \gamma_\mu) \,
{U_\mu(x)\over u_0} \, {U_\mu(x+\mu)\over u_0} \,
\psi(x+2\mu)
\nonumber \\ & &\qquad \left. 
+\bar{\psi}(x+2\mu)  \, (-{1 \over 4}\,b - {1 \over 8} \gamma_\mu) \,
{U_\mu^\dagger(x+\mu)\over u_0} \, {U_\mu^\dagger(x)\over u_0} \,
\psi(x) \right ] \, .
\label{hw}
\end{eqnarray}
The tadpole-improvement factors are explicit here.  These coefficients
remove both $O(a)$ and $O(a^2)$ errors at tree level.  The bare quark
mass is related to $\kappa$ and $b$ through
\begin{equation}
m_q={2\over 3\kappa}-4b \, .
\end{equation}
Thus the renormalized quark mass is given by
\begin{equation}
m_q={2\over 3}\left({1\over\kappa}-{1\over\kappa_{cr}}\right) \, .
\label{qmass}
\end{equation}
The wave function renormalization factor is also modified:
\begin{equation}
\psi_{\rm cont.} = \sqrt{3 \, \kappa \over 2} \, \psi_{\rm latt.}
\end{equation}
The standard Wilson action can be recovered
by truncating the two-link terms of the action, redefining $\kappa$
and renormalizing the fermion field operators.

\begin{figure}[p]
\centerline{\psfig{file=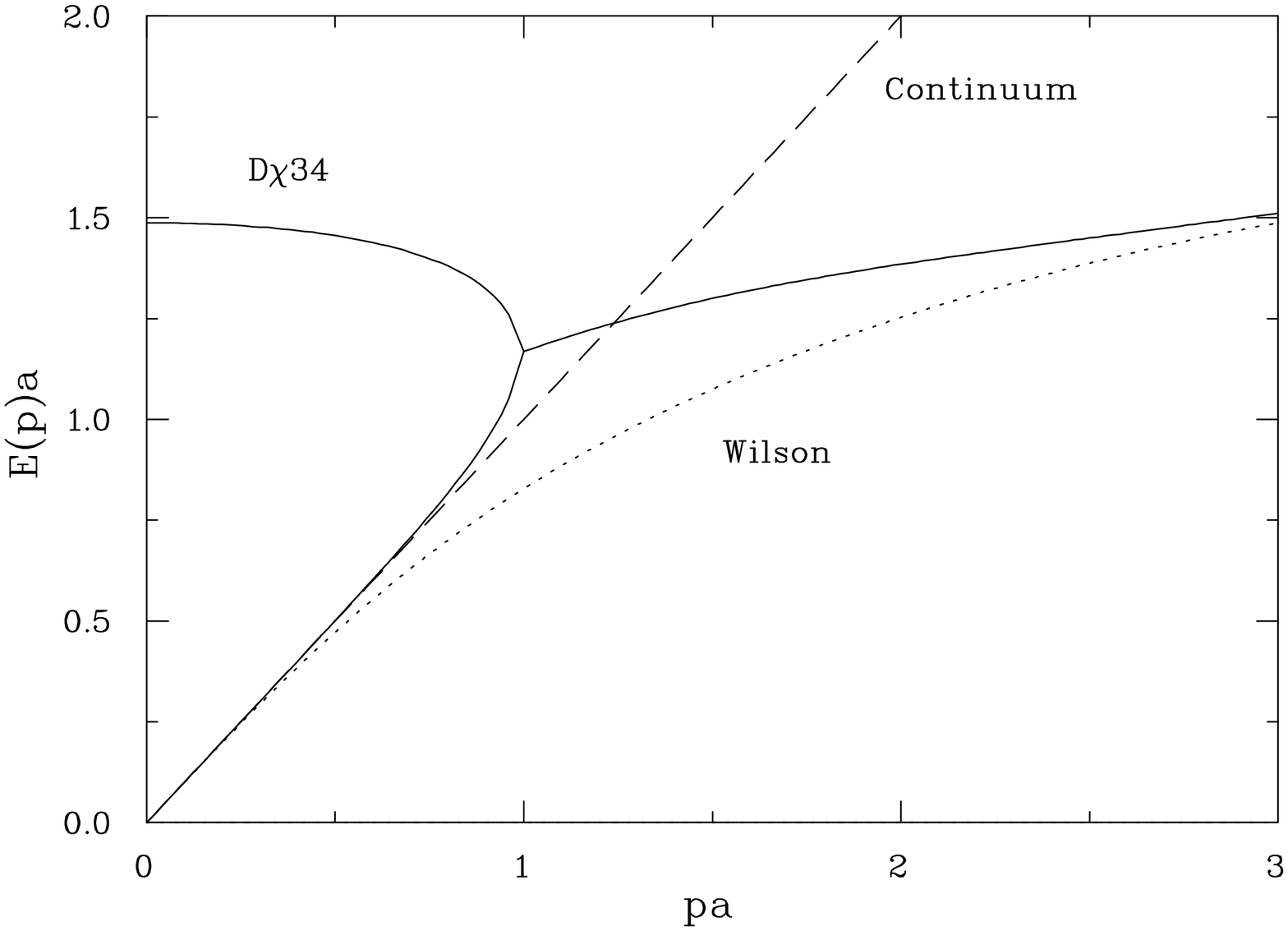,height=8cm,width=12cm,angle=90}}
\vspace{1cm}
\centerline{\psfig{file=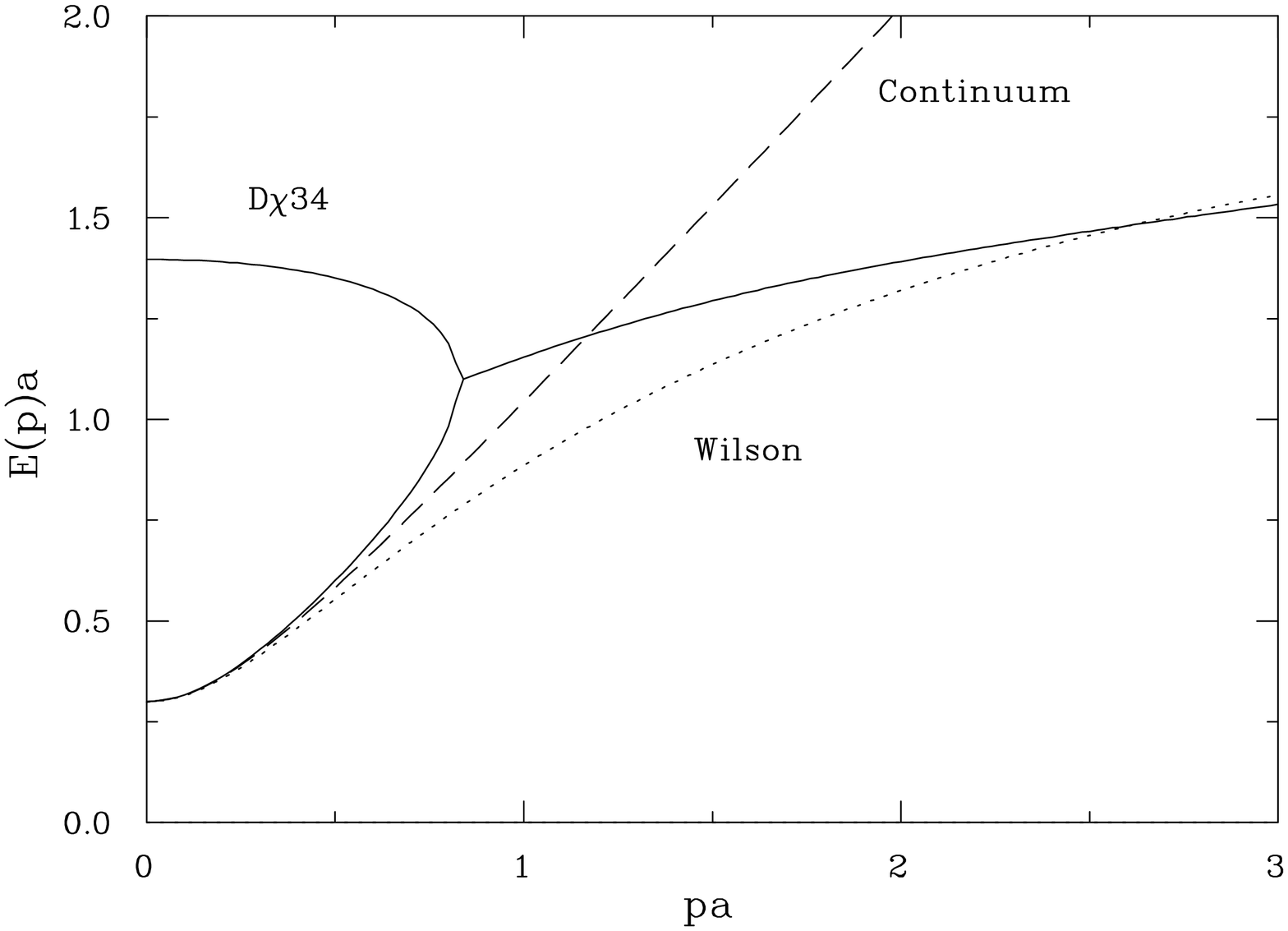,height=8cm,width=12cm,angle=90}}
\vspace{1cm}
\caption{Free dispersion relations for D$\chi34$, Wilson, and
continuum fermions.  Momentum $p$ is along the (1,1,0) direction.
Beyond the D$\chi34$ branch point, the real part of the two conjugate
roots is shown.  The upper graph is for bare mass $m \, a = 0$, the
lower $m \, a = 0.3$.}
\label{disp}
\end{figure}

Free dispersion relations can be obtained by locating the poles in the
fermion propagator.  Fig.~\ref{disp} shows free dispersion relations
for the continuum, Wilson, and D$\chi34$ fermions.  It is clear that
D$\chi34$ fermions follow the continuum more closely than Wilson
fermions.  Note that there exists an unphysical high energy doubler
(or ghost) in the D$\chi34$ action.  It is very similar to that for
the D234 action~\cite{Alf95a} and is a general feature of fermions
with next-nearest-neighbor couplings.  The doubler can be `pushed
away' from the low momentum region by various techniques, such as
tuning the value of $b$, or using an anisotropic lattice~\cite{Alf96}.
We simulate with $b = 1$, which gives good dispersion to $p \, a \sim
1$ as illustrated in Fig.~\ref{disp}.

\section{Lattice Simulations}

\subsection{Methods and Parameters}

Quenched gauge configurations were generated using the
Cabibbo-Marinari pseudo-heat-bath method~\cite{Cab82}.  Periodic
boundary conditions were used in all directions for the gauge field
and in spatial directions for the fermion field.  Dirichlet boundary
conditions were used for the fermion field in the time direction.
Configurations separated by 300 sweeps were selected after 4000
thermalization sweeps from a cold start.

\begin{figure}[t]
\centerline{\psfig{file=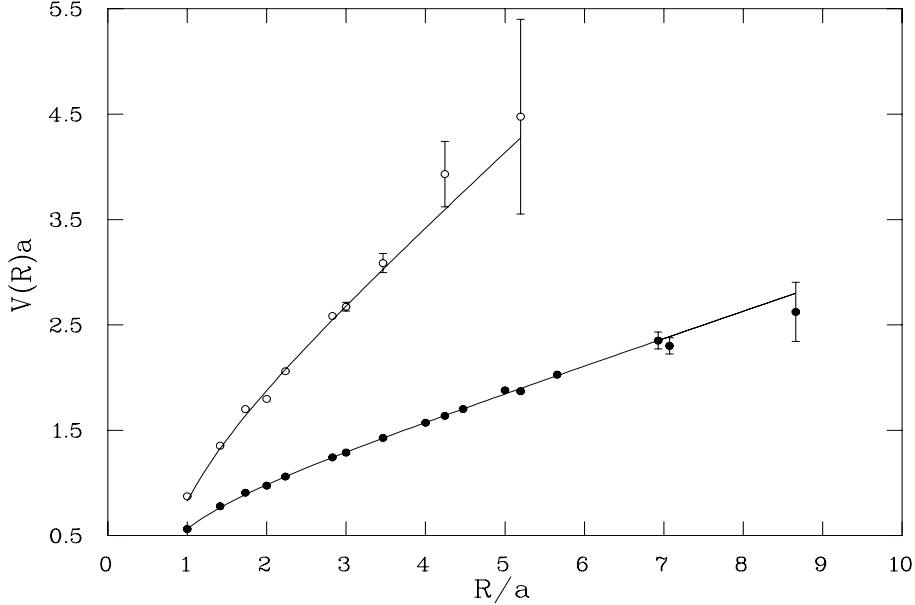,height=8cm,width=12cm,angle=90}}
\vspace{1cm}
\caption{Static potential from Wilson loops.  The empty circles are
for $\beta=6.25$, the solid circles for $\beta=7.0$.  The lines are
best fits.  The statistical errors were from 200 configurations in
both cases.}
\label{pot}
\end{figure}

Wilson loops were computed using a method developed in
Ref.~\cite{Hel94}.  Both on-axis and off-axis loops were considered.
The static potential was determined by fitting $\exp(-V(R) \, T) =
W(R,T)$.  The string tension $\sigma$ was extracted from the ansatz
$V(R) = V_0 + \sigma R - E/R$ where $V_0$ and $E$ are constants.
Fig.~\ref{pot} shows our results.  Good rotational invariance of the
static potential is observed.  Table~\ref{st} gives the extracted
string tension and the $R_0$ parameter, defined by $R^2\,dV(R)/dR=1.65
$ at $R_0$~\cite{Som94}.  Its physical value is $R_0\simeq$ 0.50 fm
and it is independent of the quenched approximation.  Using
$\sqrt{\sigma}=440$ MeV to set the scale, the lattice spacings were
determined as 0.40(6) fm and 0.23(1) fm, respectively.  Using $R_0$ to
set the scale, they are 0.40(3) fm and 0.22(1) fm.  $R_0$ depends less
on the functional form of $V(R)$ than the string tension.

\begin{table}[h]
\caption{String tension and $R_0$ extracted from the static potential.}
\label{st}
\begin{tabular}{cccccc}
$\beta$  & $\sigma a^2$ & $R_0/a$  &  $R_{min}/a$  & $R_{max}/a$  
& $\chi^2/N_{\rm DF}$ \\ \hline
6.25 & 0.81(23)  & 1.26(11)   & 1.00 & 4.24 & 1.46 \\
7.00 & 0.254(8)  & 2.279(18)  & 1.73 & 6.93 & 0.91 \\
\end{tabular}
\end{table}

Five quark propagators were computed by the Stabilized Biconjugate
Gradient algorithm~\cite{From94} for each configuration.  The five
$\kappa$ values correspond to quark masses of roughly 210, 180, 150,
120, 90 MeV, respectively, for both lattices.  The second value, 180
MeV, was taken as the strange quark mass.  A point source was used at
space-time location (x,y,z,t)=(1,1,1,2) on the $6^3\times 12$ lattice
and (1,1,1,3) on the $10^3\times 16$ lattice.  The gauge-invariant
smearing method of~\cite{Gus90} was applied at the sink to increase
the overlap of the interpolating operators with the ground states.
Table~\ref{lat} shows a summary of the parameters.

\begin{table}[b]
\caption{Summary of lattice parameters for the spectrum calculations.
$N_U$ is the number of configurations considered. 
The second kappa was selected
as the one that corresponds to the strange quark mass.}
\label{lat}
\begin{tabular}{cccccc}
Lattice & $\beta$  & $N_U$  & $u_0$  &  $\kappa$  & 
$\kappa_{cr}$   \\ \hline
$6^3\times 12$ & 6.25 & 155 & 0.820 & 
0.175, 0.177, 0.180, 0.183, 0.187 & 0.1967(2) \\
$10^3\times 16$ & 7.00 & 100 & 0.866 & 
0.170, 0.172, 0.174, 0.176, 0.178 & 0.1823(3) \\
\end{tabular}
\end{table}

Statistical errors were estimated in a third-order, single-elimination
jackknife, with bias corrections~\cite{Jack}.  A third-order jackknife
provides uncertainty estimates for the correlation functions, fits to
the correlation functions, and quantities extrapolated to the chiral
limit.

\subsection{Effective Masses}

Masses were extracted from lattice correlation functions in several
time slice intervals.  The effective hadron mass as a function of time
is defined by
\begin{equation}
M(t + 0.5) = \log[G(t)] - \log[G(t+1)]
\end{equation}
where $G(t)$ is the standard two-point function in Euclidean
space-time.  As usual, effective mass plots for the pion are very
flat.  All fits of time slices 4 through 10 on our coarse lattice and
6 through 12 on our fine lattice agree within one standard deviation.

\begin{figure}[tb]
\centerline{\psfig{file=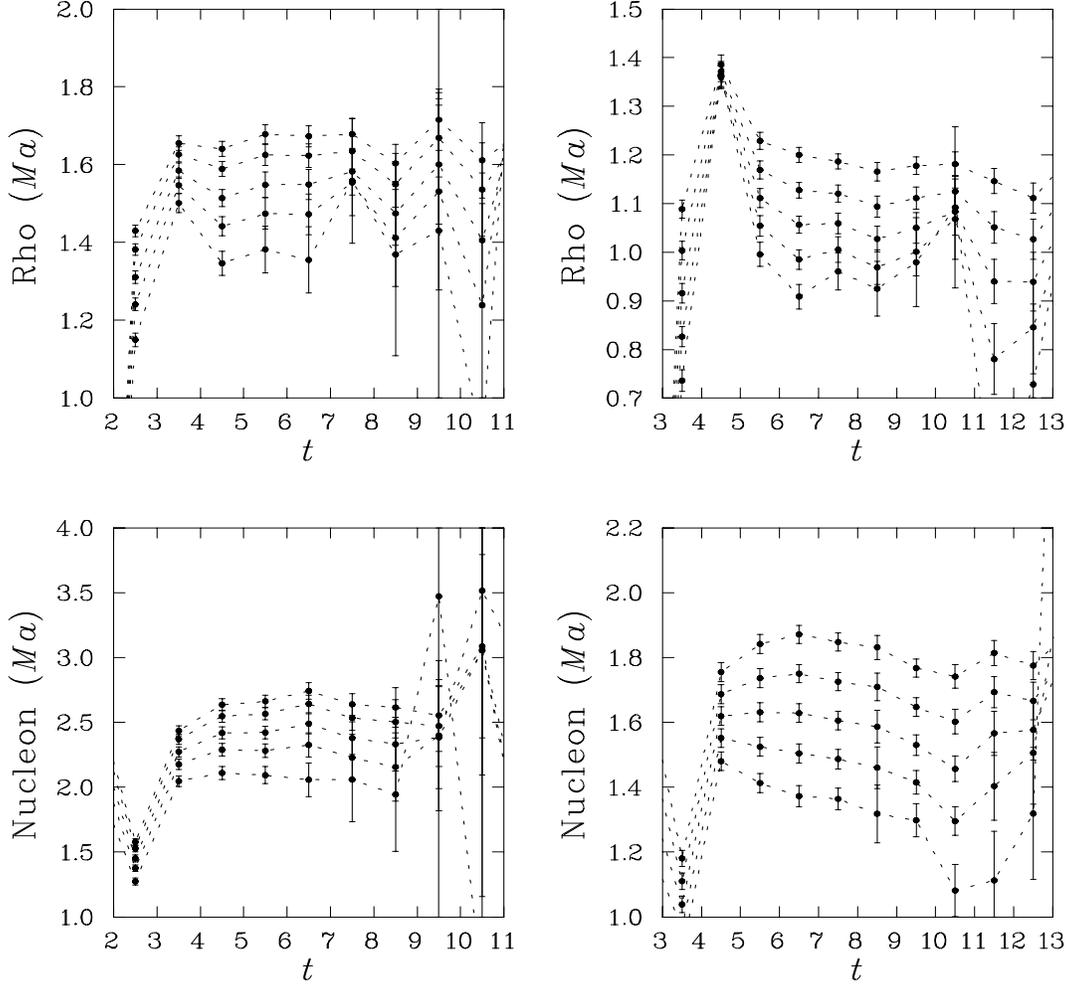,height=13cm,width=14cm}}
\vspace{1cm}
\caption{Effective mass plots for the $\rho$ and nucleon.  The first
column is for the coarser lattice, the second for the finer lattice.
The five quark masses decrease by value from top down.}
\label{Emass4}
\end{figure}

Fig.~\ref{Emass4} displays effective mass plots for the $\rho$ meson
and nucleon.  The $\rho$ meson displays good plateau behavior.  We
find $\chi^2/N_{\rm DF} \sim 0.5$ for correlation function fits from
$t = 4$ through 10 on our coarse lattice and $\chi^2/N_{\rm DF} \simeq
0.7$ for correlation function fits from $t = 6$ through 11 on our fine
lattice.  Again, all fits agree within one standard deviation.

On our coarse lattice the nucleon mass displays good plateau behavior.
The $\chi^2/N_{\rm DF}$ ranges from 1.0 to 0.1 for the lightest quark
mass.  All fits in the interval 4 through 10 agree within one standard
deviation.

On our fine lattice, plateau behavior for the nucleon mass is
excellent within the regime $t=6$ through 9.  For the heaviest three
quark masses, however, the $\chi^2/N_{\rm DF}$ identifies two separate
fitting regimes providing acceptable fits, namely 6 through 9 and 9
through 12.  The latter regime disappears for the lightest two quark
masses while the former regime grows, providing acceptable fits for
$t=6$ through 10.  Since the effective masses rise from below as a
function of $t$ for smeared sinks, we report results for the regime
$t=6$ through 9 where the signal is less likely to be dominated by
noise.

We note that this drift in the effective mass plot is only significant
as measured by the $\chi^2/N_{\rm DF}$ for the ground state nucleon
mass.  We find that other quantities such as the nucleon energy or
$\rho$-meson mass are sufficiently correlated to provide results
insensitive to the fit regime.  For example, dispersion relations
agree within one standard deviation for all fits of $t=6$ through 12
with no systematic drift in the central values.  Moreover the
extrapolated nucleon to rho mass ratios for the time slice regimes 6
through 9 and 9 through 12 are 1.44(5) and 1.41(13) respectively.
Again, the deviation of central values of the distributions is small
relative to the statistical uncertainties.  

Effective mass plots for $\Delta$ are acceptably flat but suffer a
loss of signal earlier in Euclidean time as expected.  On our coarse
lattice we find acceptable $\chi^2/N_{\rm DF} \simeq 1.0 \to 0.4$ for
intervals $t=4$ or 5 through 10 with smaller values corresponding to
lighter masses.  On our fine lattice we find acceptable $\chi^2/N_{\rm
DF} \simeq 1.0 \to 0.6$ for intervals $t=6$ or 7 through 12, again with
smaller values corresponding to lighter masses.

In summary, we report values taken from covariance matrix fits to the
time slice interval 4 through 8 on our coarse lattice and 6 through 9
on our fine lattice.  These regimes provide the best signal-to-noise
and good correlated $\chi^2/N_{\rm DF}$.

\subsection{Hadron Masses}

\begin{figure}[p]
\centerline{\psfig{file=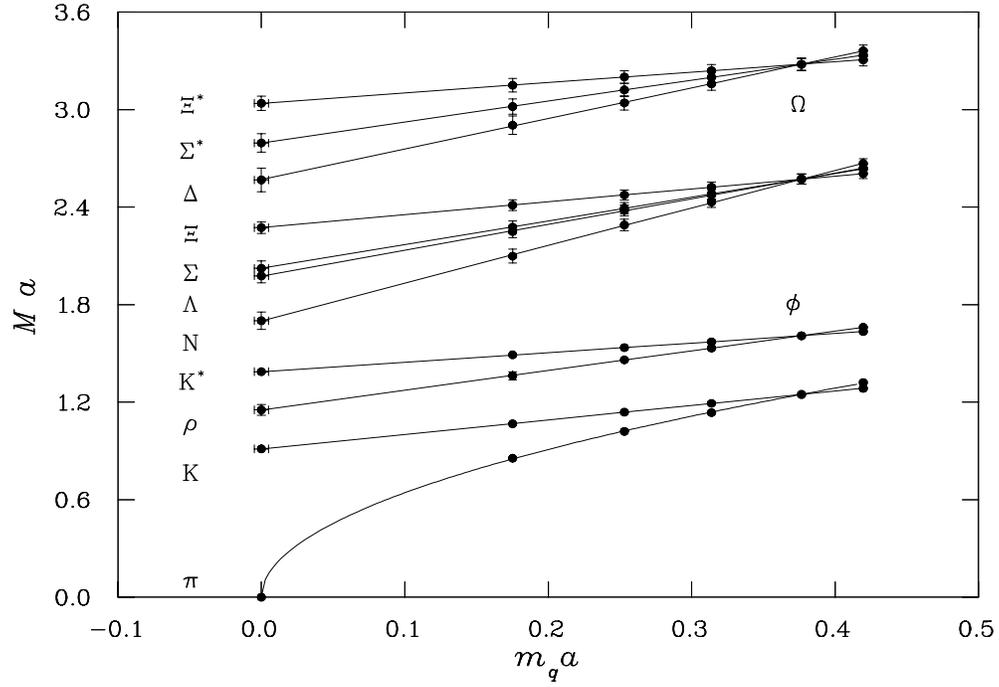,height=9cm,width=13cm,angle=90}}
\vspace{1cm}
\centerline{\psfig{file=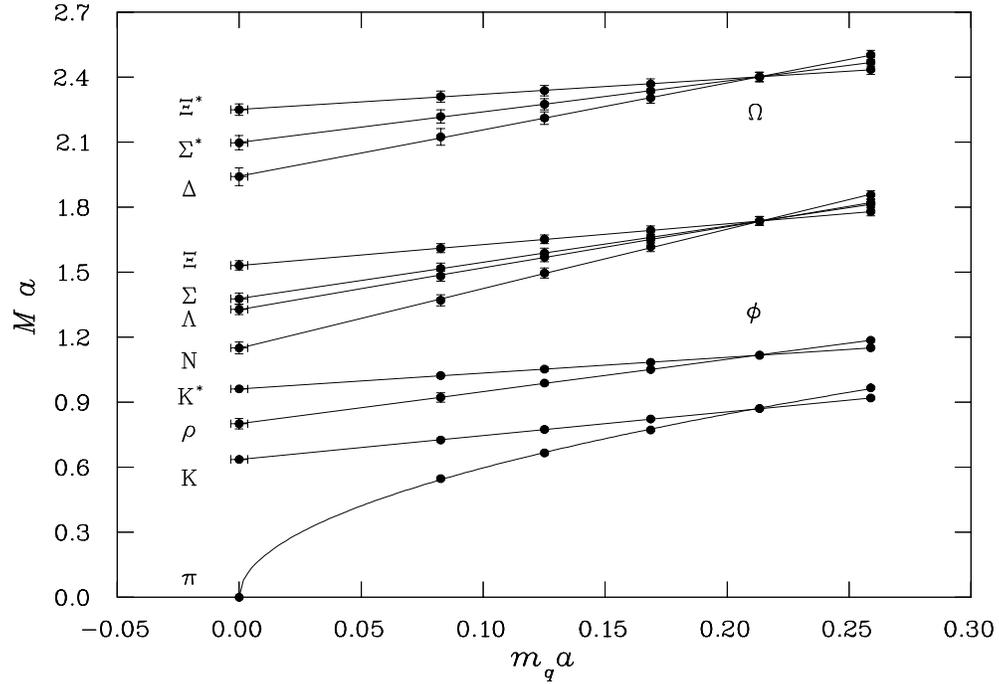,height=9cm,width=13cm,angle=90}}
\vspace{1cm}
\caption{Hadron masses in lattice units as a function of $m_q$.  The
upper graph is for the coarser lattice, the lower for the finer
lattice.  The lines are chiral fits, linear in the quark mass.  For
better viewing, the decuplet masses are shifted upward by 0.5 unit.}
\label{mall}
\end{figure}

Fig.~\ref{mall} shows the extracted hadron masses plotted as a
function of the quark mass.  For future reference, these results are
tabulated in Table~\ref{masstab}.  $\kappa_{cr}$ is determined by
linearly extrapolating $m_\pi^2$ as a function of $m_q$ to zero.  The
form $M=c_0+c_1 \, m_q$ was used for other extrapolations to the
chiral limit.  Fits with an additional term $c_2 m^{3/2}_q$ were also
considered and similar results were found with slightly larger error
bars.  The correct ordering of all the states is clearly seen at both
values of $\beta$.

\begin{table}[p]
\caption{Hadron masses as a function of $m_q$ (both in lattice units).
The results were extracted from time slices 4 to 8 on the coarser
lattice, and 6 to 9 on the finer lattice. The last column reports
results after extrapolations to the chiral limit.}
\label{masstab}
\begin{tabular}{cllllll}
\noalign{$6^3\times 12$ lattice at $\beta$=6.25}
Hadron  
& $m_q$=0.42& $m_q$=0.37& $m_q$=0.31& $m_q$=0.25& $m_q$=0.17& $m_q$=0 \\
\hline
$\pi$ 
& 1.321(6)  & 1.248(6)  & 1.136(6)  & 1.020(7)  & 0.855(8)  & 0        \\
$K$ 
& 1.29(1) & 1.25(1) & 1.19(1) & 1.14(1) & 1.07(1) & 0.91(1) \\
$\rho$
& 1.66(1) & 1.61(1) & 1.53(2) & 1.46(2) & 1.36(3) & 1.15(4)\\
$K^*$ 
& 1.64(1) & 1.61(1) & 1.57(2) & 1.54(2) & 1.49(2) & 1.39(2)\\
$N$  
& 2.67(3) & 2.57(3) & 2.43(3) & 2.29(4) & 2.10(5) & 1.70(6)\\
$\Lambda$
& 2.64(3) & 2.57(3) & 2.48(3) & 2.38(3) & 2.25(4) & 1.98(5)\\
$\Sigma$ 
& 2.64(3) & 2.57(3) & 2.48(3) & 2.39(3) & 2.28(4) & 2.02(5)\\
$\Xi$  
& 2.61(3) & 2.57(3) & 2.52(3) & 2.48(3) & 2.41(3) & 2.27(4)\\
$\Delta$
& 2.86(4) & 2.78(4) & 2.66(4) & 2.54(4) & 2.40(6)& 2.07(7)\\
$\Sigma^*$
& 2.83(4) & 2.78(4) & 2.70(4) & 2.62(4) & 2.52(5) & 2.30(6)\\
$\Xi^*$ 
& 2.82(4) & 2.78(4) & 2.74(4) & 2.70(4) & 2.65(4) & 2.54(5)\\
\\
\noalign{$10^3\times 16$ lattice at $\beta$=7.0}
Hadron  
& $m_q$=0.26& $m_q$=0.21& $m_q$=0.17& $m_q$=0.13& $m_q$=0.08& $m_q$=0 \\
\hline
$\pi$ 
& 0.967(7)  & 0.871(7)  & 0.772(7)  & 0.667(7)  & 0.548(8)  & 0        \\
$K$ 
& 0.920(7)  & 0.871(7)  & 0.823(7)  & 0.775(7)  & 0.726(7)  & 0.637(5) \\
$\rho$
& 1.19(1) & 1.12(1) & 1.05(1) & 0.99(2) & 0.92(2) & 0.80(2)\\
$K^*$ 
& 1.15(1) & 1.12(1) & 1.08(1) & 1.05(1) & 1.02(2) & 0.96(2)\\
$N$  
& 1.86(2) & 1.74(2) & 1.62(2) & 1.50(2) & 1.37(3) & 1.15(3)\\
$\Lambda$ 
& 1.82(2) & 1.74(2) & 1.65(2) & 1.57(2) & 1.48(2) & 1.33(2)\\
$\Sigma$ 
& 1.81(1) & 1.74(2) & 1.66(2) & 1.59(2) & 1.52(2) & 1.38(3)\\
$\Xi$  
& 1.78(2) & 1.74(2) & 1.69(2) & 1.65(2) & 1.61(2) & 1.53(2)\\
$\Delta$
& 2.00(2) & 1.90(2) & 1.80(2) & 1.71(3) & 1.62(4) & 1.44(4)\\
$\Sigma^*$
& 1.97(2) & 1.90(2) & 1.84(2) & 1.78(3) & 1.72(3) & 1.60(3)\\
$\Xi^*$ 
& 1.94(2) & 1.90(2) & 1.87(2) & 1.84(2) & 1.81(3) & 1.75(3)\\
\end{tabular}
\end{table}

Ratios of the chirally extrapolated masses are given in
Table~\ref{ratio} along with the ratios as observed in
nature~\cite{PDG94}.  At $\beta=7.0$ the lattice spacing estimates
follow the familiar pattern having the value based on the string
tension lying between that of the $\rho$ and nucleon based values.
This is most likely an artifact of the quenched approximation.
However, at $\beta=6.25$ we find significant disagreement among the
values and an unusual reordering of values.

\begin{table}[p]
\caption{Mass ratios after extrapolation to the chiral limit.  $a_\rho$
and $a_N$ are lattice spacings set by the rho mass (770 MeV) and the
nucleon mass (938 MeV).}
\label{ratio}
\begin{tabular}{llll}
          & $\beta$=6.25 & $\beta$=7.0  & Expt.\\
$a_{st}$  &  0.40(3) fm  & 0.220(2) fm  &      \\
$a_\rho$  &  0.30(1) fm  & 0.205(6) fm  &      \\
$a_N$     &  0.36(1) fm  & 0.242(6) fm  &      \\
\\
\noalign{Vector/Vector}
$K^*/\rho$      & 1.20(2)  & 1.20(2)  & 1.16 \\
$\phi/\rho$     & 1.40(3)  & 1.39(3)  & 1.32 \\
\\
\noalign{Octet/Octet}
$\Lambda/N$     & 1.16(2)  & 1.16(1)  & 1.19 \\
$\Sigma/N$      & 1.19(2)  & 1.20(1)  & 1.27 \\
$\Xi/N$         & 1.34(3)  & 1.33(2)  & 1.40 \\
\\
\noalign{Decuplet/Decuplet}
$\Sigma^*/\Delta$ & 1.12(1)  & 1.11(1)  & 1.12 \\
$\Xi^*/\Delta$    & 1.24(3)  & 1.22(2)  & 1.24 \\
$\Omega/\Delta$   & 1.34(3)  & 1.33(2)  & 1.36 \\
\\
\noalign{Pseudoscalar/Vector}
$K/\rho$        & 0.79(2)  & 0.79(2)  & 0.64 \\
\\
\noalign{Octet/Vector}
$N/\rho$        & 1.48(6)  & 1.44(5)  & 1.22 \\
$\Lambda/\rho$  & 1.72(6)  & 1.66(5)  & 1.45 \\
$\Sigma/\rho$   & 1.76(6)  & 1.72(5)  & 1.55 \\
$\Xi/\rho$      & 1.97(6)  & 1.91(6)  & 1.71 \\
\\
\noalign{Decuplet/Vector}
$\Delta/\rho$   & 1.79(7)  & 1.77(5)  & 1.60 \\
$\Sigma^*/\rho$ & 2.00(6)  & 1.97(6)  & 1.80 \\
$\Xi^*/\rho$    & 2.20(7)  & 2.16(6)  & 1.99 \\
$\Omega/\rho$   & 2.40(7)  & 2.35(6)  & 2.17 \\
\\
\noalign{Decuplet/Octet}
$\Delta/N$      & 1.21(5)  & 1.23(2)  & 1.31 \\
$\Sigma^*/N$    & 1.35(5)  & 1.37(2)  & 1.47 \\
$\Xi^*/N$       & 1.49(4)  & 1.50(3)  & 1.63 \\
$\Omega/N$      & 1.63(4)  & 1.64(3)  & 1.78 
\end{tabular}
\end{table}

Focusing first on ratios of hadrons having the same angular momentum,
we see very little change in the values as the lattice spacing is
decreased.  These ratios are also remarkably similar to those observed
in nature, despite the fact that these are quenched QCD calculations.
In addition, these ratios support our selection for the strange quark
mass.

This close resemblance to nature is not shared by ratios of hadrons
with different angular momentum.  All four classes of ratios
significantly disagree with those of nature.  Once again we see the
familiar quenched artifact of the Octet/Vector ratio being too large
and the Decuplet/Octet ratio being too small.  

\begin{figure}[tb]
\centerline{\psfig{file=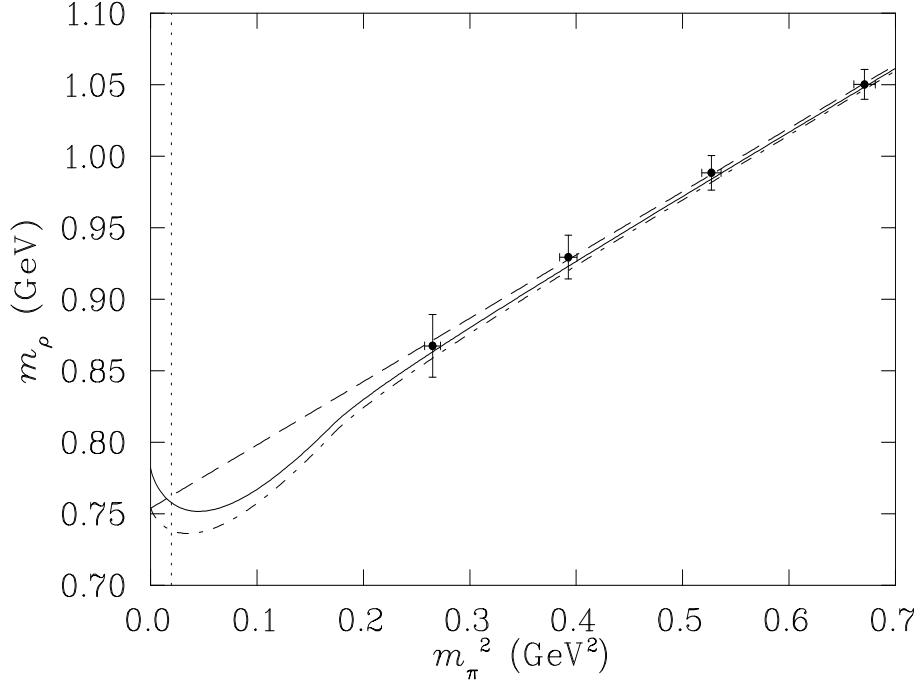,height=9cm,width=12cm,angle=90}}
\vspace{1cm}
\caption{Plot of the $\rho$-meson mass as a function of the squared
pion mass obtained from our finer lattice.  $a_\rho$ has been used to
set the scale.  The dashed line illustrates the standard linear
extrapolations of $m_\pi^2$ and $m_\rho$.  The solid and dot-dash
curves include the two-pion self-energy of the $\rho$
meson~\protect\cite{Lein94} for dipole dispersion cut-off values of 1
and 2 GeV respectively.  The increase in the slope at $m_\rho/m_\pi =
1.8$ ($m_\pi^2 \simeq 0.21$) provided by the two-pion self energy is
the right order of magnitude to restore agreement with the empirical
value.}
\label{Jgraph}
\end{figure}

The standard failure of the $K/\rho$ mass ratio in the quenched
approximation is also seen here.  This shortcoming has been widely
realized through an examination of the $J$-parameter~\cite{Lac95}
defined by
\begin{equation}
J = m_\rho \, {dm_\rho \over dm_\pi^2} \Bigm |_{m_\rho/m_\pi = 1.8} 
\;\simeq\; m_{K^*} \, {m_{K^*} - m_\rho \over m_{K}^2 - m_\pi^2 } \, .
\end{equation}
Empirically this ratio is 0.48.  However we find 0.40 on our coarse
lattice and 0.38 on our fine lattice.  The physics associated with
this discrepancy was first reported in Ref.~\cite{Lein94}.  There it
was pointed out that the self-energy generated by two-pion
intermediate states of the $\rho$-meson, which is excluded in the
quenched approximation, acts to increase the $J$ parameter.  
Fig.~\ref{Jgraph} provides a sketch of how including the two-pion
self-energy of the $\rho$ can increase the value of $J$ from 0.38 to
0.46.

Perhaps the most important information displayed in Table~\ref{ratio}
is that the Octet/Vector mass ratios display less than satisfactory
scaling for the larger lattice spacing.  To further examine scaling
and make contact with other studies, we focus on the the $N/\rho$ mass
ratio which is among the the most revealing of ratios.

\begin{figure}[p]
\centerline{\psfig{file=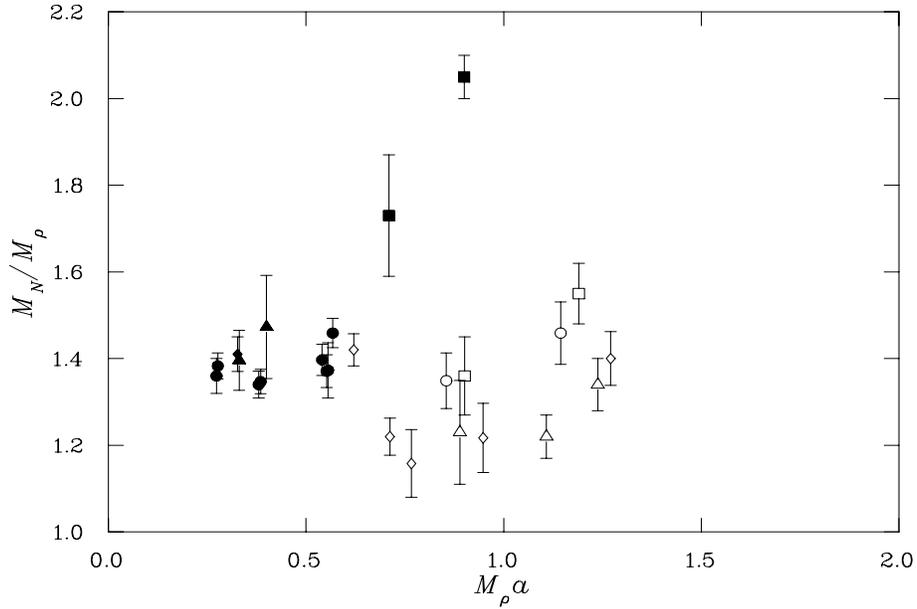,height=8cm,width=12cm,angle=90}}
\vspace{0.5cm}
\caption{The $N/\rho$ mass ratio versus $M_\rho a$ at the chiral limit.
Solid symbols denote the standard Wilson action.  Open symbols denote
improved actions including SW \protect\cite{Col97} $(\diamond)$, D234
\protect\cite{Alf95a} $(\triangle)$, D$\chi34$ \protect\cite{Fie96}
$(\Box)$, and D$\chi34$ $(\circ)$ (this work).}
\label{nrho}
\end{figure}
\begin{figure}[p]
\centerline{\psfig{file=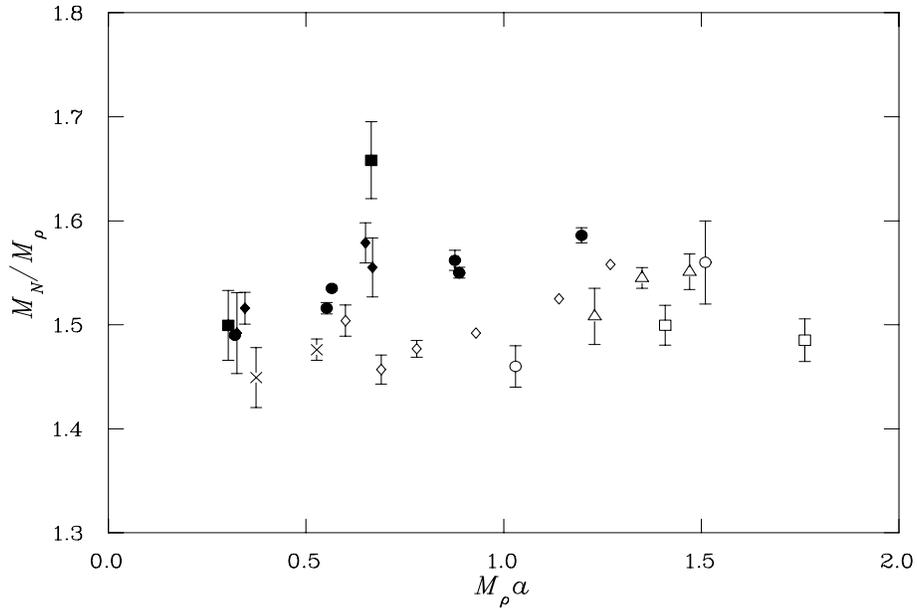,height=8cm,width=12cm,angle=90}}
\vspace{0.5cm}
\caption{The $N/\rho$ mass ratio versus $M_\rho a$ at a fixed
$\pi/\rho$ mass ratio of 0.7 for various actions.  The solid symbols
denote the standard actions: Wilson (square and diamond), staggered
(circle).  The open symbols denote improved actions:
nonperturbatively-improved SW~\protect\cite{Goc97} $(\times)$,
fixed-point action~\protect\cite{DeG97} $(\Box)$, SW $(\diamond)$,
D234 $(\triangle)$, and D$\chi34$ $(\circ)$ (this work). }
\label{nrho07}
\end{figure}

Fig.~\ref{nrho} shows a comparison of the $N/\rho$ mass ratio versus
$M_\rho \, a$ at the chiral limit.  Fig.~\ref{nrho07} shows the
$N/\rho$ mass ratio as a function of $M_\rho \, a$ at a fixed
$\pi/\rho$ mass ratio of 0.7~\cite{DeG97b}.  This method is free of
complications from chiral extrapolations.  Both cases clearly show the
improvement provided by the D$\chi34$ action.  Indeed, the D$\chi34$
action has reproduced the state-of-the-art quenched QCD ratios using
unimproved actions at coarse lattice spacings of 0.24 fm.  Our results
at the chiral limit are compatible with those in~\cite{Fie96}, but lie
slightly above the SW and D234 results.  The results from various
improved actions at fixed $\pi/\rho$ ratio show a certain degree of
universality over a wide range of lattice spacings.  The performance
of fixed-point actions at very coarse lattice spacings is remarkable.

\subsection{Dispersion and Rotational Symmetry}

In addition to mass ratios, hadron states at finite momentum
projections $\vec{p} \, a =\vec{n}(2\pi/L)$ were also calculated.
Dispersion was examined by calculating the effective speed of light,
$c_{\scriptscriptstyle eff}$, defined by $c_{\scriptscriptstyle
eff}^2=(E^2(p)-E^2(0))/p^2$, which is to be compared with 1.

A comparison with SW and D234 lattice actions~\cite{Alf95a} is made in
Table \ref{dispComp}.  The lattice spacings are based on charmonium
for SW and D234 actions whereas the static quark potential is used for
our results.  As such, the lattice spacings in Table \ref{dispComp}
are approximate.  Lattice volumes range from 2.0 to 2.4 fm.  D$\chi34$
results are based on simulations at $\kappa = 0.183$ for our coarse
lattice and $\kappa = 0.176$ on our fine lattice.  The dispersion for
the $O(a)$-improved SW action is very poor relative to the excellent
dispersions of the next-nearest-neighbor improved D$x34$ actions.  The
D$\chi34$ dispersion is excellent even at our coarse lattice spacing.

\begin{table}[thb]
\caption{Comparison of SW, D234 and D$\chi34$ actions for the speed of
light squared obtained from the dispersion of $\pi$ and $\rho$ mesons
at $m_\pi/m_\rho \simeq 0.7$ for $p \, a=(2\pi/L)$.}
\label{dispComp}
\begin{tabular}{rcccc}
Hadron &$a$ (fm) &SW         &D234         &D$\chi34$  \\
\hline
$\pi$  &0.40     &0.63(2)    &0.95(2)      &0.99(3)    \\
$\pi$  &0.24     &           &0.99(4)      &1.04(4)    \\
\hline
$\rho$ &0.40     &0.48(4)    &0.93(3)      &0.93(6)    \\
$\rho$ &0.24     &           &1.00(6)      &0.99(6)    \\
\end{tabular}
\end{table}

Rotational symmetry is explored in Table~\ref{dispt}, which reports
results for $m_q \sim 180$ MeV.  At the coarser lattice spacing, some
drift in the central values is seen for the pion and nucleon.  The
drift in the pion is similar to that seen for the D234c action
reported in Ref.~\cite{Lep97}.  However, the drift in dispersion for
the $\phi$ meson reported in Ref.~\cite{Lep97} is not apparent in our
results for the D$\chi34$ action.  However, a similar drift may be
hidden in the uncertainties.  The D$\chi34$ action has much better
rotational symmetry than the SW action\cite{Lep97}.  Moreover, the
D$\chi34$ action provides satisfactory dispersion at our finer lattice
spacing and is competitive with the D234 action\cite{Alf95a}.

\begin{table}[thb]
\caption{Evaluation of dispersion and rotational invariance via the
effective speed of light.  The results are for $m_q\sim$ 180 MeV.}
\label{dispt}
\begin{tabular}{cccccc}
$a$ (fm) & $\vec{n}$& $\pi$    & $\phi$   & $N$     & $\Omega$  \\
\hline
0.40	 & (1,0,0)  & 0.98(2)  & 0.91(4)  & 1.00(7)  & 0.99(12) \\
	 & (1,1,0)  & 0.91(4)  & 0.91(6)  & 0.94(8)  & 0.91(7)  \\
	 & (1,1,1)  & 0.86(9)  & 0.92(10) & 0.92(7)  & 0.90(10) \\
\hline
0.24	 & (1,0,0)  & 1.04(3)  & 1.02(4)  & 1.10(6)  & 1.06(7) \\
	 & (1,1,0)  & 1.05(4)  & 1.02(4)  & 1.06(5)  & 1.11(5) \\
	 & (1,1,1)  & 0.98(6)  & 0.98(6)  & 1.06(6)  & 1.06(5) \\ 
\end{tabular}
\end{table}

\section{Conclusion}

We have computed masses and dispersion relations of light hadrons in
lattice QCD using tree-level $O(a^2)$ tadpole-improved gauge and
fermion actions.  Compared to standard Wilson actions, the gluon
action has an additional six-link (rectangle) term while the D$\chi34$
fermion action has an additional two-link (next-nearest-neighbor)
term.  These actions have the appeal of being simple to implement and
inexpensive to simulate.

A great deal of effort is being directed toward finding the ultimate
improved action that will facilitate simulations on the coarsest of
lattices.  We note however, that many quantities of phenomenological
interest such as hadron form factors involve momenta on the order of a
GeV.  As such, a highly improved action which is costly to simulate
may not be the ideal action for hadron phenomenology, especially for
exploratory purposes.

The mass ratios obtained from the D$\chi34$ action at 0.24 fm on a
modest $10^3 \times 16$ lattice reproduce the state-of-the-art results
using conventional unimproved actions.  Excellent dispersion and
rotational invariance up to $pa\approx 1$ are also found.  These
results demonstrate that the D$\chi34$ action can serve as a viable
candidate for the study of hadron phenomenology, and in our view is
preferable to the highly-improved but more costly D234 action.  We
plan to use the D$\chi34$ action to study hadron properties beyond the
spectrum, such as multipole form factors of hadrons in general.  These
results also bode well for future explorations beyond the quenched
approximation.

\acknowledgements 
It is a pleasure to thank R. Woloshyn and H. Trottier for helpful
discussions in the early stage of this work, and T. DeGrand and
T. Kovacs for their help in computing the static potential.  We also
thank T. DeGrand for a critical reading of this manuscript.
Simulations on our smaller lattice were performed on a DEC Alpha
workstation at TRIUMF.  Simulations on our larger lattice were
performed on the Cray C90 at NERSC through resources awarded to
Institute of Nuclear Theory.  This work was supported in part by the
U.S. DOE under Grants DE-FG03-93DR-40774 and DE-FG06-88ER40427, and by
the Natural Sciences and Engineering Research Council of Canada and
the Australian Research Council.

\end{document}